\documentstyle[aps,pre,amssymb,multicol]{revtex}
\tighten
\draft
\newtheorem{lemma}{Lemma}
\newtheorem{thrm}{Theorem}
\begin{document}
\title{Classifying Rational Densities Using Two One-Dimensional Cellular
 Automata}
\author{H.~F. Chau\footnote{e-mail address: hfchau@hkusua.hku.hk}, K.~K. Yan,
 K.~Y. Wan and L.~W. Siu}
\address{Department of Physics, University of Hong Kong, Pokfulam Road, Hong
 Kong}
\date{\today}
\preprint{HKUPHYS-HFC-05}
\maketitle
\begin{abstract}
 Given a (finite) string of zeros and ones, we report a way to determine if the
 number of ones is less than, greater than, or equal to a prescribed number by
 applying two sets of cellular automaton rules in succession. Thus, we solve
 the general density classification problem using cellular automaton.
\end{abstract}
\begin{multicols}{2}
\section{Introduction}
\label{S:Intro}
 Cellular automaton (CA) is an extremely simple local interaction model of
 natural systems \cite{Wolfram}. Although the rules of CA are simple and local,
 complicated spatial patterns can be formed. Moreover, from the computer
 science point of view, CA can be regarded as a very special kind of Turing
 machine with no internal memory. In fact, tailor-made CA can be used to
 simulate certain logical operations \cite{Korotkov}. Therefore, it is
 instructive to know the power and limitations of CA in computation.
\par
 Since CA is essentially a special Turing machine with no internal memory, it
 is natural to ask if CA can be used to perform certain task that requires a
 global counter. An example is the so-called density classification problem:
 Consider a one-dimensional (finite) chain of sites with periodic boundary
 condition. Each site is either in state zero or state one. The problem is to
 change the state of every site to zero if the number of zeros is more than
 that of ones in the one-dimensional chain. Otherwise, every site is set to
 state one. The density classification problem is trivial if one can invoke an
 internal memory to count the number of zeros in the chain. On the other hand,
 Land and Belew showed that density classification cannot be done perfectly
 using one-dimensional CA \cite{No-go}.
\par
 Later on, Fuk\'{s} pointed out that the density classification problem can be
 solved if we apply {\em two} CA rules in succession \cite{r1/2}. For a
 one-dimensional lattice of $N$ sites, he first applies the Wolfram's
 elementary CA rule~184 (the so-called traffic rule \cite{1d-bml})
 $\left\lfloor (N - 2)/2 \right\rfloor$ times. Then, he applies the Wolfram's
 elementary CA rule~232 (the so-called majority rule) $\left\lfloor (N - 1)/2
 \right\rfloor$ times. The combined automaton solves the density classification
 problem succinctly \cite{r1/2}.
\par
 With slight modifications, Fuk\'{s} is able to classify density in the form
 $1/n$ for some integer $n\geq 2$ \cite{r1/2}. Then, Fuk\'{s} went on to ask if
 it is possible to use combined CA to classify an arbitrary density $\rho_c$.
 Here, we report a simple solution to his problem using a combination of two
 CAs provided that $\rho_c$ is a rational number. But before stating our CA
 rules, let us re-state the density classification problem in the most general
 context.
\par
 Consider a one-dimensional lattice making up of $N$ sites with periodic
 boundary condition. The state of each site can either be zero or one. One is
 given a critical density $\rho_c$ between zero and one. The initial density
 of ones in the system is denoted by $\rho$. (That is, there are $N\rho$ sites
 in the system with states equal one.) Our goal is to find a combination of two
 CA rules such that after going through the CA, the final states of the sites
 on the lattice becomes all zeros (and all ones) if $\rho < \rho_c$ (and $\rho
 > \rho_c$, respectively). And in the case of $\rho = \rho_c$, the final state
 consists of exactly $N\rho_c$ sites with state equals one.\footnote{From the
 non-linear dynamics point of view, the all zeros and all ones are the two
 stable fixed points of the CA dynamics, whereas the density equals $\rho_c$ is
 an unstable fixed point of the CA dynamics.}
\par
 Since the number of sites $N$ is finite, it suffices for us to consider the
 case when $\rho_c$ is a rational number written in the form $p/q$ for some
 relatively prime integers $p$ and $q$. Furthermore, we can also assume that
 $p \neq 0$ and $p\neq q$ since classifying $\rho_c = 0$ and $1$ are trivial
 using CA. Now, we report the two CA rules we use to solve the density
 classification problem for rational critical density.
\section{Modified Traffic Rule}
\label{S:Traffic}
 Our generalized traffic rule goes as follows: First, we regard a site with
 state one as being occupied by a car. Otherwise, that site is empty. Then a
 car can move in the next timestep to the right by one site if and only if (a)
 its immediate right-neighboring-site is unoccupied; and (b) the $q-1$
 consecutive right-neighboring-sites is occupied by at most $p-1$ cars.
 Otherwise, the car stays in its original position. For example, if $\rho_c =
 2/5$, then the first cars in the sequences $10000$ and $10100$ can move one
 step forward in the next timestep, while the first cars in the sequences
 $11000$ and $10011$ cannot. Readers can readily verify that the total number
 of cars in the system is conserved under the above set of rules. Moreover, one
 can easily convert the above set of traffic rules into a {\em finite} CA rule
 table consisting of $2^{q+1}$ rules. In addition, when $\rho_c = 1/2$, the
 modified traffic rule is reduced to Wolfram's elementary CA rule~184.
\par
 We define the local car density at each site to be the total number of cars
 contained in that site and in its $q-1$ immediate right-neighboring-sites
 divided by $q$. Thus, the local car density of a site changes when and only
 when a car enters that site, or a car leaves the $(q-1)$th site to its right.
 For simplicity, a collection of sites is said to be in the low density region
 if the local car density of each site in the collection is less than or equal
 to $\rho_c$. And a high density region is defined as a collection of sites
 with local car densities greater than $\rho_c$.
\par
 Now, we prove two theorems concerning the distribution of high and low density
 regions under the repeated actions of the modified traffic rules.
\begin{lemma}~
 If the initial density $\rho$ of the system is less than or equal to $\rho_c$,
 then after at most $\left\lceil N (\max (q, 2p) -1) / q \right\rceil + q-2$
 timesteps, the local car density for every site is less than or equal to
 $\rho_c$. \label{Lemma:1}
\end{lemma}
\par\smallskip\noindent
{\em Proof:} Consider a collection of sites with local car density greater than
 $\rho_c$ in the initial system configuration. Using the modified traffic rule,
 in the co-moving frame of a car, no site will have car density exceeding
 $(p+1)/q$ if that car is originally located at a site with local car density
 less than or equal to $\rho_c$. In addition, such an aggregation of car above
 the density threshold $\rho_c$ will be achieved within the first $q-1$
 timesteps. Since the overall car density of the system is less than or equal
 to $\rho_c$, so upon repeated applications of the modified traffic rules, cars
 will gradually move out of local regions with car density exceeding $\rho_c$
 (if any). Note that the local car densities of those cars moving out of these
 ``high density regions'' are less than or equal to $\rho_c$. Moreover, once
 these cars are ``dissolved'' from a local high density region, the local car
 density of the sites containing these cars will never exceed $\rho_c$ unless
 these cars merge into a high car density region in front.
\par
 Suppose none of the cars moving out of a high car density region will be
 stopped by another high car density region in front. Then, once all the high
 car density regions dissolve completely, no further high density region will
 be formed thereafter. Thus, our assertion that all sites will have local car
 density not exceeding $\rho_c$ asymptotically is true in this case. So, it
 remains to consider the case that some of the cars moving out of a high car
 density region move into (and hence temporarily stopped by) a high car density
 region in front. Since the overall car density of the entire system does not
 exceed $\rho_c$, at least one of the high car density regions will start to
 dissolve. Note that $N$ is finite, and those cars merging into a high density
 region will readily re-dissolve again once they are allowed to move. After a
 finite number of timesteps, all the high density regions will disappear. After
 that, no more high density region can form. Thus, our assertion that all sites
 will have local car density not exceeding $\rho_c$ is also true is this case.
\par
 Finally, we estimate the number of timesteps required to reach this state.
 Since the updating is taken in parallel, the worst case occurs when some of
 the local high density regions are formed during the first $q-2$ timesteps.
 Then all the cars merge into a single high car density region before they
 finally dissolve. According to the modified traffic rules, if $p\leq q/2$,
 then all cars dissolved from a high density region will not be blocked. In
 this case, it takes at most $q-1$ timesteps to dissolve $p$ cars from a high
 density region. On the other hand, if $p > q/2$, then some of the dissolved
 cars will still be blocked occasionally. And it takes $q-1+(2p-q) = 2p-1$
 timesteps to dissolve $p$ cars from a high density region. Therefore, in any
 case, it requires at most $\left\lceil N \rho (\max (q, 2p) -1) / p
 \right\rceil \leq \left\lceil N (\max (q, 2p) -1) / q \right\rceil$ timesteps
 to completely dissolve a high density region making up of $N\rho$ cars. Hence,
 our assertion is proved.
\hfill$\Box$
\begin{lemma}~
 If the initial density $\rho$ of the system is greater than $\rho_c$, then
 after at most $\left\lceil N (q-p) (\max (q,2p) - 1) / pq \right\rceil$
 timesteps, the local car density for every site is greater than or equal to
 $\rho_c$. \label{Lemma:2}
\end{lemma}
\par\smallskip\noindent
{\em Proof:} We may assume that the local car density at each site initially
 is less than or equal to $\rho_c$. Otherwise, no car can move right at the
 beginning and our assertion is trivially true. Since the overall car density
 $\rho$ is greater than $\rho_c$, all the low car density regions (sites with
 local car density less than $\rho_c$) are surrounded by high car density ones
 (sites with car density greater than or equal to $\rho_c$). Thus, cars can
 gradually move from the high density to the low density regions. And using
 similar arguments as in Lemma~\ref{Lemma:1}, after finite number of timesteps,
 the entire system is contained in a single high density region. And from that
 time on, no car in the system can move.
\par
 We move on to estimate the number of timesteps required to reach this
 ``frozen'' state. Similar to the argument in Lemma~\ref{Lemma:1}, the worst
 case occurs when there is only one low density region initially. And it is
 easy to verify that the number of timesteps required to reach a frozen state
 for such an initial system configuration equals $\left\lceil N (1 - \rho)
 (\max (q, 2p) - 1) / p \right\rceil \leq \left\lceil N (q-p) (\max (q,2p) -1)
 / pq \right\rceil$.
\hfill$\Box$
\par\bigskip\indent
 Combining Lemmas~\ref{Lemma:1} and~\ref{Lemma:2}, we conclude that
\begin{thrm}~
 By applying the modified traffic rules for $\left\lceil N (\max (q,2p) - 1)
 \max (q-p,p) / pq \right\rceil + q-2$ times, an initial system configuration
 will be segregated into one of the following three cases:
\par
 (a) If $\rho < \rho_c$, then the local car density at every site is less than
 or equal to $\rho_c$. In addition, at least one of the sites will have local
 car density strictly less than $\rho_c$.
\par
 (b) If $\rho = \rho_c$, then the local car density at every site equals
 $\rho_c$.
\par
 (c) If $\rho > \rho_c$, then the local car density at every site is greater
 than or equal to $\rho_c$. In addition, at least one of the sites will have
 local car density strictly greater than $\rho_c$. \label{Th:1}
\end{thrm}
\section{Modified Majority Rule}
\label{S:Majority}
 After segregating the system configuration according to its initial density
 $\rho$, the density classification problem becomes straight-forward. We
 consider the following modified majority rule for a given critical density
 $\rho_c \equiv p/q$: The state of a site in the next timestep is one if there
 are at least $2p+1$ ones in the $2q+1$ sites consisting of itself and the $q$
 consecutive left- and right-neighboring-sites. Otherwise, the state of this
 site in the next timestep is zero. For example, if $\rho_c = 1/2$, then the
 states of the middle sites in the next timestep for $10101$ and $01010$ are
 one and zero, respectively.
\par
 Now, we present the results of applying the modified majority rule to certain
 system configurations which are of interest.
\begin{lemma}~
 Any system configuration with local car density at every site equals $\rho_c$
 is a fixed point of the modified majority rule dynamics. \label{Lemma:3}
\end{lemma}
\par\smallskip\noindent
{\em Proof:} Consider an arbitrary site $\alpha$ in the system. Since the local
 car density at every site equals $\rho_c$, there are precisely $2p$ sites in
 state one among the $2q$ neighboring sites of $\alpha$. Thus, the majority
 rule implies that the state of site $\alpha$ in the next timestep is equal to
 its present state. Consequently, this system configuration is a fixed point of
 the modified majority rule.
\hfill$\Box$
\begin{lemma}~
 Suppose the local car density at every site is greater than or equal to
 $\rho_c$ and also that the total density of the system is strictly greater
 than $\rho_c$. After applying the modified traffic rule $\left\lceil N / 2
 (q-1) \right\rceil$ times, the state of every site in the system will become
 one. Similarly, if the local car density at every site is less than or equal
 to $\rho_c$ and also that the total density of the system is strictly less
 than $\rho_c$. Then after applying the modified traffic rule $\left\lceil N /
 2 (q-1) \right\rceil$ times, the state of every site in the system will become
 zero. \label{Lemma:4}
\end{lemma}
\par\smallskip\noindent
{\em Proof:} For simplicity, we only consider the case that $\rho > \rho_c$.
 The proof for the case that $\rho < \rho_c$ is similar. From
 Lemma~\ref{Lemma:3}, we know that the local car density at every site must be
 greater than or equal to $\rho_c$ after repeated applications of the modified
 majority rule. Moreover, if the local car density of a site $\alpha$ exceeds
 $\rho_c$, then the states of $\alpha$ and its $q-1$ left- and
 right-neighboring-sites must be one in the next (and hence all subsequent)
 timestep. In other words, the propagation speed for state one is $q-1$ sites
 per timestep both leftward and rightward. Therefore, after $\left\lceil N / 2
 (q-1) \right\rceil$ timesteps, all sites in the system will be in state one.
 Hence, our assertion is proved.
\hfill$\Box$
\par\bigskip\indent
 Combining Theorem~\ref{Th:1}, Lemmas~\ref{Lemma:3} and~\ref{Lemma:4}, we
 obtain the following CA density classification theorem:
\begin{thrm}[Density Classification By CA]~
 Let $\rho_c = p/q$ be a rational number between zero and one, with $p,q$ being
 relatively prime positive integers. Then the density classification problem
 can be solved using the following two CA rules: apply the modified traffic
 rules $\left\lceil N (\max (q,2p) - 1) \max (q-p,p) / pq \right\rceil + q-2$
 times and then followed by the modified majority rule $\left\lceil N / 2 (q-1)
 \right\rceil$ times. \label{Th:2}
\end{thrm}
\section{Discussions}
\label{S:Diss}
 In summary, we report a way to classify the density of ones using two
 one-dimensional binary CAs provided that the density threshold $\rho_c$ is a
 rational number. Our result, therefore, generalizes that of Fuk\'{s}
 \cite{r1/2}. Besides, our construction also takes care of the case when the
 density $\rho$ is equal to its critical value $\rho_c$ --- something Fuk\'{s}
 does not consider seriously. Because of the local nature of CA rules, we
 believe that the time complexity of a general density classification problem
 using any combination of CA rules is at least ${\mathrm O} (N)$. Thus, apart
 from a constant speedup, our set of density classification CA rules is
 probably the least time-consuming.
\par
 As we have discussed in Sec.~\ref{S:Intro}, if the number of site in the
 system $N$ is finite, it suffices to restrict ourselves to consider rational
 density classification. Any irrational density can be well approximated by a
 corresponding rational density. Actually, to approximate an irrational density
 by some better and better rational numbers, the size of the CA rule table
 grows. Thus, we believe that combinations of CA rules is unlikely to be
 powerful enough to classify irrational density in an infinite one-dimensional
 system.
\acknowledgements
 This work is supported by the University of Hong Kong CRCG grant under the
 contract number 335/025/0040.

\end{multicols}
\end{document}